\documentclass[runningheads]{llncs}

\usepackage[T1]{fontenc}

\usepackage{latexsym}
\usepackage{amsfonts}
\usepackage{graphicx}
\usepackage{hyperref}
\usepackage{xcolor}
\usepackage{multirow}

\newcommand{\Z}{\mathbb{Z}}

\newcommand{\G}{\mathbb{G}}

\begin{document}

\title{BASS: \lowercase{\uppercase{B}oolean \uppercase{A}utomorphisms \uppercase{S}ignature \uppercase{S}cheme}}

\author{Dima Grigoriev \inst{1}  \and Ilia Ilmer \inst{2}  \and Alexey Ovchinnikov  \inst{3}  \and Vladimir Shpilrain \inst{4}}

\authorrunning{Grigoriev et al.}

\institute{CNRS, Math\'ematiques, Universit\'e de Lille, 59655, Villeneuve d'Ascq, France \email{Dmitry.Grigoryev@univ-lille.fr} \and Department of Computer Science, CUNY Graduate Center, 365 5th Ave, New York, NY 10016 \email{i.ilmer@icloud.com} \and Department of Mathematics, Queens College, City University of New York, Queens, NY 11367 \email{alexey.ovchinnikov@qc.cuny.edu} \and Department of Mathematics, The City College of New York, New York,
NY 10031 \email{shpilrain@yahoo.com}}

\maketitle

\begin{abstract}
We offer a digital signature scheme using Boolean automorphisms of a multivariate  polynomial algebra over integers. Verification part of this scheme is based on the approximation of the number of zeros of a multivariate Boolean function.

\end{abstract}

\section{Introduction}

Due to the concern that if large-scale quantum computers are ever built, they will compromise the security of many commonly used cryptographic algorithms, NIST had begun in 2016 a process to develop new cryptography standards and, in particular, solicited proposals for new digital signature schemes \cite{NIST} resistant to attacks by known quantum algorithms, such as e.g. Shor's algorithm \cite{Shor}. In particular, there is an interest in signature schemes whose security is based on new assumptions.

One possible way to avoid quantum attacks based on solving the hidden subgroup problem (including the attacks in \cite{Shor}) is {\it not} to use one-way functions that utilize one or another (semi)group structure. The candidate one-way function that we use in our scheme here takes a private polynomial automorphism $\varphi$ as the input and outputs $\varphi(P)$ for a public multivariate polynomial $P$.

To avoid any parallels with the encryption scheme of \cite{Moh}, we say up front that since ours is just a signature scheme (i.e., is not a spin-off of any encryption scheme), we do not need our candidate one-way function to have a trapdoor because the private key holder does not need to invert the function. Also, in \cite{Moh}, the candidate one-way function was $\varphi$ itself, and the private (decryption) key was $\varphi^{-1}$. In contrast, in our signature scheme $\varphi^{-1}$ does not play any role and does not have to be computed.

The main novelty of our signature scheme is manifested in the verification part. First, note that any polynomial $P$ has as many zeros as $\varphi(P)$ does, where $\varphi$ is any automorphism of the polynomial algebra. To balance between security and efficiency, we do not want the number of zeros to be either too small or too large. To that end, we use polynomials over integers, but we count zero values on Boolean tuples only. Since the number of Boolean tuples is exponential in the number of variables, it can still be too large to process deterministically. Instead, we use a non-deterministic (Monte Carlo) method to estimate the number of zero (or nonzero) values of a polynomial in question. We note that the accuracy of the Monte Carlo method for estimating the number of zeros of a multivariate polynomial was studied and quantified in \cite{Karpinski}.

\section{Scheme description}\label{description}

Let $K=\Z[x_1, \ldots, x_n]$ denote the algebra of polynomials in $n$ variables over the ring $\Z$ of integers, and let $B(K)$ denote the factor algebra of $K$ by the ideal generated by all  polynomials of the form $(x_i^2-x_i)$, $i=1, \ldots, n$. Informally, one can call $B(K)$ the ``Booleanization" of $K$. We note that the ring $B(K)$ is isomorphic (as a ring) to the direct sum of $2^n$  copies of the ring $\Z$.
\medskip

\noindent The signature scheme is as follows.
\medskip

\noindent {\bf Private:}  an automorphism $\varphi$ of the algebra $B(K)$. We note that $\varphi$ is defined by the polynomials $y_i=\varphi(x_i), ~i=1,\ldots, n$.
\medskip

\noindent {\bf Public:}

\noindent -- 3 sparse polynomials $P_i=P_i(x_1, \ldots, x_n)$ with coefficients $\pm 1$.\\
-- 3 polynomials $\varphi(P_i)$, where $\varphi$ is a private automorphism of the algebra $B(K)$. We note that $\varphi(P_i)=P_i(y_1, \ldots, y_n)$, where $y_i=\varphi(x_i)$.\\
-- a hash function $H$ with values in the algebra $B(K)$ and a (deterministic) procedure for converting values of $H$ to sparse polynomials from the algebra $B(K)$.\\
-- a set $\G$ of polynomials. This set includes, in particular, all monomials and all polynomials of the form (1-monomial). See Section \ref{automorphism}) for more details.
\medskip

\begin{remark}
We emphasize that the automorphism $\varphi$, the 3 sparse polynomials $P_i$, and the 3 polynomials $\varphi(P_i)$ are all generated/computed in the offline phase. The hash function $H$ is one of the standard hash functions (we suggest SHA3-256), with values converted to a polynomial in  $B(K)$ (see Section \ref{hash}).

\end{remark}

\medskip

\noindent {\bf Signing} a message $m$:
\medskip

\noindent {\bf 1.} Apply the hash function $H$ to the message $m$. Convert $H(m)$ to a polynomial $Q=Q(x_1, \ldots, x_{n+1})$ with integer coefficients using a deterministic public procedure (see Section \ref{hash}). That is, the polynomial $Q$ has an extra variable compared to the polynomials $P_i$.
\medskip

\noindent {\bf 2.} The automorphism $\varphi$ is extended to the ``Booleanization" of the algebra $\Z[x_1, \ldots, x_{n+1}]$ by taking $x_{n+1}$ to $x_{n+1}+r(x_1, \ldots, x_n) -2x_{n+1} \cdot r(x_1, \ldots, x_n)$, where $r(x_1, \ldots, x_n)$ is a random  polynomial from the set $\G$ of polynomials (see Section \ref{automorphism}). This extended automorphism we denote by the same letter $\varphi$.
(The fact that this is, indeed, an automorphism of the ``Booleanization" is part of Proposition \ref{positive} in Section \ref{automorphism}.)

\medskip

\noindent {\bf 3.} The signature is $\varphi(Q)$.
%\medskip
\smallskip

\begin{remark} The reason why we extend the automorphism $\varphi$ by adding an extra variable $x_{n+1}$ at Step 2 is to prevent the forger from accumulating many pairs $(Q, \varphi(Q))$ with the same $\varphi$. Now, with each new signature, we have a different $\varphi$ because of a random choice of the polynomial $r(x_1, \ldots, x_n)$ at Step 2.
\end{remark}

\medskip

\noindent {\bf Verification:}

\medskip

\noindent {\bf 1.} The verifier computes $H(m)$ and converts $H(m)$ to $Q=Q(x_1, \ldots, x_{n+1})$ using a deterministic public procedure.

\medskip

\noindent {\bf 2.} The verifier selects a random 4-variable  polynomial $u(x,y,z,t)$ from $B(\Z[x,y,z,t])$ with coefficients 0, 1, -1, 2, or -2, and computes $u(\varphi(P_1), \varphi(P_2), \varphi(P_3), \varphi(Q))$. Note that this is equal to $\varphi(u(P_1, P_2, P_3,Q))$. Denote the polynomial $\varphi(u(P_1, P_2, P_3,Q))$ by $S$.
\medskip

\noindent {\bf 3.} The verifier also computes $u(P_1, P_2, P_3, Q)$. Denote this polynomial by $R$. (Note that $S$ should be equal to $\varphi(R)$ if the signature is valid.)
\medskip

\noindent {\bf 4.} The verifier then compares the proportion of positive values on Boolean tuples for the polynomials $R$ and $S$. That is, the proportion of positive values on $(n+1)$-tuples $(x_1, \ldots, x_{n+1})$, where each $x_i$ is 0 or 1. These proportions are estimated using a non-deterministic (Monte Carlo) method. 

The verifier accepts the signature if and only if these proportions for $R$ and $S$ are different in no more than 3\% of the total number of trials in the Monte Carlo method.
(See Section \ref{automorphism} for an explanation of why these proportions should be exactly the same when computed deterministically if $S$ is an automorphic image of $R$.)
\medskip

\begin{remark} 
With suggested parameters, the number of Boolean $(n+1)$-tuples is quite large ($2^{n+1}$, to be exact). Given that counting zeros (or non-zeros) on Boolean tuples is \#P-hard, see \cite{Valiant}, it is computationally hard to count the number of positive values on Boolean tuples {\it precisely}, which is why the verifier has to use a non-deterministic method. We explain the method in the following subsection. 
\end{remark}
\vskip 0.1 cm

\noindent {\bf Correctness.}
While it is obvious that polynomials $P$ and $\varphi(P)$ have the same number of zeros, it is not at all obvious why they have the same number of positive values on Boolean tuples. Indeed, this may not be true for an arbitrary automorphism $\varphi$, so we have a special algorithm for sampling $\varphi$. This is explained in Section \ref{automorphism}, and correctness is formally proved in the Appendix.

\subsection{Monte Carlo method for counting positive values of a polynomial on Boolean tuples}

Our non-deterministic method for estimating the proportion of positive values on Boolean tuples for a given polynomial $P$ is pretty standard. Just plug in a large number of random Boolean tuples into $P$ and count how many of them yield a positive value of $P$. Then divide the obtained number by the total number of Boolean tuples used; this is your proportion.

We note that the accuracy of the Monte Carlo method for counting {\it zeros} of Boolean polynomials was studied and quantified in \cite{Karpinski}. See our Section \ref{accuracy} for more details on the accuracy.

%How to sample a random two-variable  polynomial $u(x,y)$ with coefficients from $\F_2$?

\section{Key generation}\label{generation}

First we note that, since the algebra $B(K)$ is the factor algebra of $K$ by the ideal generated by all polynomials of the form $(x_i^2-x_i)$ and since we only count values of a polynomial on Boolean tuples,  when we generate the public polynomials $P_i$ it makes sense to only generate monomials where no $x_j$ occurs with an exponent higher  than 1. Then generating $P_i$ will look as follows.

\subsection{Generating a random $t$-sparse polynomial}\label{polynomial}

\begin{enumerate}

\item  Select, uniformly at random, an integer $d$ between 1 and $b$ (where $b$ is one of the parameters of the scheme). This integer will be the degree of our monomial. (Note that the degree of a monomial cannot be higher  than $n$ since our monomials are square-free because of factoring by the ideal generated by all polynomials of the form $(x_i^2-x_i)$.)
\medskip

\item  To select a monomial of degree $d$,  do a selection of $x_i$, uniformly at random from $\{x_1, \ldots, x_n\}$, $d$ times, avoiding repetition of $x_i$. Then build the monomial as a product of the selected $x_i$.
\medskip

\item  Finally, build a $t$-sparse polynomial as a linear combination of $t$ selected  monomials with coefficients $\pm 1$, selected at random.
\end{enumerate}

\subsection{Generating a random polynomial from the set $\G$}\label{G}

The set $\G$ of polynomials in $\Z[x_1, \ldots, x_n]$ will play a crucial role in generating automorphisms of the algebra $B(K)$, see Section \ref{automorphism}. This set
can be defined recursively as follows. Assign all variables $x_1, \ldots, x_n$ to $\G$. Then keep adding more polynomials to $\G$ using the following rules: (1) if a polynomial $P$ belongs to $\G$, then $1-P$ belongs to $\G$, too; (2) if both polynomials $P_1$ and $P_2$ belong to $\G$, then their product $P_1P_2$ belongs to $\G$, too.

\begin{remark} The number of multiplications in the above procedure for generating a polynomial from the set $\G$ (see Step 3 in the procedure below) is one of the parameters of our scheme; denote it by $r$.
\end{remark}

Note that the set $\G$ consists of polynomials $P$ such that $P(x_1, \ldots, x_n)=$  0 or 1 for any Boolean $n$-tuple $(x_1, \ldots, x_n)$. This easily follows by induction  from the above recursive definition of the set $\G$. In other words, any polynomial from $\G$ induces an $n$-variable Boolean function and, conversely, any $n$-variable Boolean function is induced by a polynomial from $\G$.

Based on this description, we suggest the following procedure for sampling a polynomial, depending on variables from a subset $X$ of the set of variables, from the set $\G$. We emphasize again that in our scheme, this is done in the offline phase.

\begin{enumerate}

\item Select a random monomial as in the previous Section \ref{polynomial}, except that the degree $d$ should be really small,  1 or 2. Denote this monomial by $M$. %and the monomial should not depend on $x_k$.
\medskip

\item With probability $\frac{1}{2}$, select between $M$ and $1-M$. Denote the result by $M'$.
\medskip

\item Select, uniformly at random, a variable $x_i$ not from the subset $X$ of variables. Then, with probability $\frac{1}{2}$, multiply $M'$ by either $x_i$ or $1-x_i$.
\medskip

\item  Repeat steps (2) through (3) $r$ times for some small $r$ (one of the parameters of the scheme).

\end{enumerate}

\subsection{Converting $H(m)$ to a polynomial}\label{hash}

We suggest using a hash function $H$ from the SHA-3 family, specifically SHA3-256. We assume the security properties of SHA3-256, including collision resistance and preimage resistance. Below is an ad hoc procedure for converting a hash $H(m)$ to a polynomial. We assume there is a standard way to convert $H(m)$ to a bit string of length 256.

Let $B$ be a bit string of length 256. We will convert $B$ to a polynomial from the factor algebra of $K=\Z[x_1, \ldots, x_{n+1}]$ by the ideal generated by all polynomials of the form $(x_i^2-x_i)$, $i=1, \ldots, n+1$. We note that this process is deterministic.

\medskip

\noindent {\bf (1)} Split 256 bits in 32 8-bit blocks.  The 5 leftmost bits will be responsible for a coefficient of the corresponding  monomial, while the 3 rightmost bits will be responsible for a collection of variables $x_i$ in the monomial.
\medskip

\noindent {\bf (2)} After Step (1), we have 32 3-bit blocks corresponding to monomials of degree 3 that we now have to populate with 3 variables each. Enumerate 96 bits in these 32 3-bit blocks by $x_1, \ldots, x_{32}, x_1, \ldots, x_{32}, x_1, \ldots, x_{32}$ (in this order, going left to right). Now each 3-bit block is converted to a monomial that is a product of $x_i$ corresponding to the places in the bit string where the bit is ``1". In particular, each monomial will be of degree at most 3.
\medskip

\noindent {\bf (3)} Now we have to use 5 remaining bits in each 8-bit block to obtain
an integer coefficient for each monomial of degree $\le 3$ obtained at Step 2. This is done as follows. First, we compute the sum of these 5 bits. Then, we reduce it modulo 3.
If the result is 0, then the coefficient is 0. If the result is 1, then the coefficient is 1. If the result is 2, then the coefficient is -1.
\medskip

\noindent {\bf (4)} Combine all monomials and coefficients obtained at Steps (2), (3)  into a polynomial.

\section{Generating an automorphism $\varphi$}\label{automorphism}

An automorphism $\varphi$ is generated offline, as follows.

Recall that the set $\G$ consists of polynomials $P$ such that $P(x_1, \ldots, x_n)=$  0 or 1 for any Boolean $n$-tuple $(x_1, \ldots, x_n)$, see Section \ref{G}.

Then we have:

\begin{proposition}\label{positive}
Let $h=h(x_1,\ldots, x_n)$ be a polynomial from the set $\G$. Suppose $h$ does not depend on $x_k$.
Let $\alpha$ be the map that takes $x_k$ to $x_k + h - 2x_k \cdot h$ and fixes all other variables.
Then:\\
{\bf (a)} $\alpha$ defines an automorphism of $B(K)$, the factor algebra of the algebra $\Z[x_1, \ldots, x_n]$ by the ideal generated by all polynomials of the form $(x_i^2-x_i)$, $i=1, \ldots, n$. Denote this automorphism also by $\alpha$.
\medskip

\noindent {\bf (b)} The group of automorphisms of $B(K)$ is generated by all automorphisms as in part (a) and is isomorphic to the group of permutations of the vertices of the $n$-dimensional Boolean cube.
\medskip

\noindent {\bf (c)} For any polynomial $P$ from $\Z[x_1, \ldots, x_n]$, the number of positive values of $P$ on Boolean tuples $(x_1, \ldots, x_{n})$ equals that of $\alpha(P)$.
\end{proposition}

\noindent For the proof of Proposition \ref{positive}, see the Appendix.

\subsection{Generating triangular automorphisms}\label{triangular}

Our (private) automorphism $\varphi$ will be a composition of ``triangular" automorphisms and permutations on the set of variables. Below is how we generate an ``upper triangular" automorphism $\alpha$.

\medskip

\noindent {\bf (1)} Let $k=1$.
%Choose an index $k$ at random between 1 and $n$.
\medskip

\noindent {\bf (2)} With probability $\frac{1}{2}$, either take $x_k$ to itself or take $x_k$ to $x_k + h(x_1,\ldots, x_n) - 2x_k \cdot h(x_1,\ldots, x_n)$, where $h(x_1,\ldots, x_n)$ is a random $t$-sparse polynomial from the set $\G$ not depending on any $x_j$ with $j \le k$ (see Section \ref{G}). Fix all other variables.
\medskip

%\noindent {\bf Comment.} The polynomial $h(x_1,\ldots, x_n)$ should be ``small", e.g. $h(x_1,\ldots, x_n)=x_i + x_j^2 + x_rx_s$.
%\medskip

\noindent {\bf (3)} If $k<n$, increase $k$ by 1 and go to Step (2). Otherwise, stop.
%Apply a random permutation to the set $x_1,\ldots, x_n$ of variables.
\medskip

Generating a ``lower triangular" automorphism $\beta$ is similar:

\medskip

\noindent {\bf (1)} Let $k=n$.
\medskip

\noindent {\bf (2)} With probability $\frac{1}{2}$, either take $x_k$ to itself or take $x_k$ to $x_k + h(x_1,\ldots, x_n) - 2x_k \cdot h(x_1,\ldots, x_n)$, where $h(x_1,\ldots, x_n)$ is a random $t$-sparse polynomial from the set $\G$ not depending on $x_j$ with $j \ge k$. Fix all other variables.
\medskip

\noindent {\bf (3)} If $k>1$, decrease $k$ by 1 and go to Step (2). Otherwise, stop.

\subsection{Generating $\varphi$ as a composition of triangular automorphisms and permutations}\label{composition}

Having generated an upper triangular automorphism $\alpha$ and a lower triangular  automorphism $\beta$, we generate our private automorphism $\varphi$ as a composition
$\alpha \beta \pi$, where $\pi$ is a  random permutation on the set of variables. Here $\alpha$ is applied first, followed by $\beta$, followed by $\pi$.

At the end of the whole procedure, we will have $n$ polynomials $y_i=\varphi(x_i)$ that define the automorphism $\varphi$.

\section{Suggested parameters}\label{parameters}

\noindent For the hash function $H$, we suggest SHA3-256.

\noindent For the number $n$ of variables, we suggest $n = 31$.

\noindent For the number $t$ of monomials in $t$-sparse polynomials, we suggest $t=3$.

%\noindent For the degree $d$ of monomials in $t$-sparse polynomials, we suggest $d=3$.
%\noindent For the parameter $m$ at Step (4) in Section \ref{automorphism}, we suggest $m=n$.

\noindent For the bound $b$ on the degree of monomials in $t$-sparse polynomials,  we suggest $b=3$.

\noindent For the degree $d$ of the monomial $M$ in the procedure for generating a polynomial from the set $\G$ (Section \ref{G}), we suggest $d=2$.

\noindent For the number $r$ of the number of multiplications in the procedure for generating a polynomial from the set $\G$ (Section \ref{G}), we suggest $r=1$.

\noindent For the number of trials in Monte Carlo method for counting positive values of a polynomial on Boolean tuples, we suggest 3,000.

\section{Performance and signature size} \label{performance}

For our computer simulations, we used Apple MacBook Pro, M1 CPU (8 Cores), 16 GB RAM computer.
Julia code is available, see \cite{Julia}.

With the suggested parameters, signature verification takes about 0.3 sec on average, which is not bad, but the polynomial $\varphi(Q)$ (the signature) is rather large, almost 4 Kb on average.

The size of the private key (the automorphism $\varphi$) is about 1.5 Kb, and the size of the public key is about 12.5 Kb.

We note that we have measured the size of a signature, as well as the size of private/public keys, as follows.
We have counted the total number of variables that occurred in relevant polynomial(s) and multiplied that number by 5, the number of bits sufficient to describe the index of any variable (except $x_{32}$). To that, we added the number of monomials times 3 (the average number of bits needed to describe a coefficient at a monomial in our construction(s).

%The size of $\varphi$ itself (i.e., the total number of monomials in all the polynomials $\varphi(x_i)$) is about 1500 monomials on average.

As usual, there is a trade-off between the size of the private key $\varphi$ and its security. The size of $\varphi$ can be reduced to just a few hundred monomials, but then security becomes a concern since some of $\varphi(x_i)$ may be possible to recover more or less by inspection of the public pairs $(P_i, \varphi(P_i))$.

In the table below, we have summarized performance data for most reasonable (in our opinion) parameter sets. Most columns are self-explanatory; the last column shows memory usage during verification.

\bigskip

\hskip -1cm\begin{tabular}{|p{1.3cm}|p{1.3cm}|p{1.3cm}|p{1.5cm}|p{1.6cm}|p{1.3cm}|p{1.4cm}|p{1.4cm}| p{1.3cm}| }
 \hline
 \multicolumn{9}{|c|}{Performance metrics for various parameter values} \\
 \hline

   \# monomials in $P_i$ & max degree of $P_i$ & max degree of mono- mials~$M$
   & parameter $r$ & verification time (sec) & signature size (Kbytes) & public key size (Kbytes)& private key size (Kbytes) & memory usage (Mbytes)  \\
 \hline
    3   & 3 & 1& 1& 0.3& 4.3 & 17.5 & 1.2 & 5.7 \\
  \hline

    3   & 3 & 2& 1& 0.3& 3.7 & 12.6 & 1.6 & 5.7 \\
 \hline
  3   & 4 & 1& 1& 0.5& 4.3 & 25 & 1.2 & 5.8 \\

 \hline
  4  & 3   & 1 & 1& 4.1& 4.2 & 38 & 1.25 & 7.1 \\
 \hline
   5 & 3&  1& 1& 6.2& 6 & 46 & 1.3 & 8.2 \\
 \hline
   3 & 3&  1& 2& 2& 20 & 56 & 5.5 & 6.3\\
 \hline

\end{tabular}

\bigskip

\subsection{Accuracy of the Monte Carlo method}\label{accuracy}

We have run numerous computer simulations to estimate the probability of a ``false positive" result, in particular accepting a forged signature from somebody who knows only some of $\varphi(x_i)$. In our experiments, the difference between the number of positive values of $u$ and $u'$ for a $u'$ obtained by using a wrong private key $\varphi$ was always above 9\%. Recall that the threshold difference for accepting a signature in our scheme is 3\%.

``False negative" results (i.e., rejecting a valid signature because the difference was more than 3\%) are not as critical as ``false positive" results are, but it is still better to avoid them.  Increasing the number of trials in the Monte Carlo method obviously reduces the probability of false negative (as well as false positive) results. To quantify this statement, one can use the formula from \cite[Theorem 1]{Karpinski}:

\begin{equation}\label{1}
N \ge C \cdot \frac{4 \log(\frac{2}{\delta})}{\epsilon^2}
\end{equation}

for some constant $C$. Here $\delta$ is the probability that the Monte Carlo method gives a wrong answer, and $\epsilon$ is the accuracy we want. (In our case,  $\epsilon = 3\% = 0.03$.) Then, $N$ is the number of trials needed to provide the desired accuracy with the desired probability. %Without going into details of how the coefficient $\mu$ is defined, we just say that in our situation, $\mu$ is approximately 0.5, so the formula (\ref{1}) becomes

According to our computer simulations, in $1000$ trials there is one false negative result on average. This suggests that the constant $C$ in our situation is about  0.02.

%\begin{equation}\label{2}
%N \ge  \frac{8 \log(\frac{2}{\delta})}{\epsilon^2}.
%\end{equation}

Therefore, with the recommended 3000 trials the probability of a false negative result will be about $2^{-33}$.

Thus, it is not surprising that with 3000 trials, we did not detect any false negative or false positive results in any of our computer simulations. %However, the verification time then increases to 1.2 sec on average.

%\bigskip

\section{What is the hard problem here?}\label{problem}

Recall that the candidate one-way function that we use in our scheme takes a private polynomial automorphism $\varphi$ as the input and outputs $\varphi(P)$ for a public multivariate polynomial $P$. Thus, the (allegedly) hard problem here is: given a public pair (or several pairs) $(P, \varphi(P))$, recover $\varphi$. We note that such a $\varphi$ does not have to be unique, although most of the time it is.

The problem of recovering $\varphi$ from a pair $(P, \varphi(P))$, as well as the relevant decision problem to find out whether or not, for a given pair of polynomials $(P, Q)$, there is an automorphism that takes $P$ to $Q$, was successfully addressed only for two-variable polynomials \cite{Makar}. For polynomials in more than two variables the problem is unapproachable at this time, and there are no even partial results in this direction. This is, in part, due to the fact that there is no reasonable description of the group of automorphisms of $\Z[x_1, \ldots, x_n]$ when $n>2$, so even a ``brute force" approach based on enumerating all automorphisms is inapplicable.

Of course, in a cryptographic context one is typically looking not for general  theoretical results, but rather for practical ad hoc, often non-deterministic, attacks. The most straightforward non-deterministic attack that comes to mind here is as follows. Recall that monomials in the polynomial $P$ have low degree (bounded by 3). Thus, given a monomial, say, $x_1x_2x_3$ in the polynomial $P$, one can try to replace each $x_i$ by a hypothetical $\varphi(x_i)$ of the form $\sum (c_i x_i +c_{ij}x_ix_j + c_{ijk}x_ix_jx_k)$, with indeterminate coefficients $c_i, c_{ij}, c_{ijk}$. Given that $\varphi$ is ``sparse", this may yield a number of equations in the indeterminate coefficients that is not huge. However, these equations will include not just linear equations, but also equations of degree 2 and 3 (since $\varphi(x_1x_2x_3)=\varphi(x_1)\varphi(x_2)\varphi(x_3)$), and given a large number (hundreds) of unknowns $c_i, c_{ij}, c_{ijk}$, there is no computationally feasible way known to solve such a system.

In the next Section \ref{attack}, we offer a ``linearization" of this attack where all equations become linear, at the expense of making the number of unknowns and the number of equations very large.

\section{Linear algebra attack}\label{attack}

One can attempt to recover the private automorphism $\varphi$ from the public pairs $(P_i, \varphi(P_i))$ by using linear algebra, more specifically by trying to replace $\varphi$ by a linear transformation of the linear space of monomials involved in $P_i$ and in the polynomials $\varphi(x_i)$. The latter polynomials are not known to the adversary, but at least the degrees of monomials in those polynomials can be bounded based on the public polynomials $\varphi(P_i)$.

Let us compute the dimension of the linear space of monomials of degree at most 27 in 31 variables. This is because a polynomial $P_i$ has monomials of degree at most 3, and in the polynomials $\varphi(x_i)$ there can be monomials of degree up to 9 (with the suggested parameters), so in $\varphi(P_i)$ there can be monomials of degree up to 27.

By a well-known formula of counting combinations with repetitions, the number of monomials of degree at most 27 in 31 variables is equal to ${57}\choose{30}$ $\approx 1.4 \cdot 10^{16} > 2^{53}$. This is how many variables the attacker will have should (s)he use a linear algebra attack. The number of equations will be about triple of this number.

Solving a system of linear equations with that many variables and equations would require more than $2^{53 \cdot 2.3} \approx 2^{122}$ arithmetic operations, according to our understanding of the state-of-the-art in solving systems of linear equations.

We note that increasing the number of variables in the polynomial algebra will not seriously affect efficiency as long as the bound on the degrees of monomials remains the same. At the same time, the more variables the less feasible the linear algebra attack is.

\section{Security claims} \label{Security}

The linear algebra ``brute force" attack amounts to solving a system of linear equations (over $\Z$) with about $2^{53}$ variables and at least as many equations.

There could be ad hoc attacks on the public key aiming at recovering some of the $\varphi(x_i)$, but recovering only some of $\varphi(x_i)$ does not make the probability of passing verification non-negligible, according to our computer simulations.

We have not been able to come up with any meaningful ideas of forgery without getting a hold of the private key.

As for quantum security, we do not make any general claims, just mention that since there are no abelian (semi)groups in play in our scheme, Shor's quantum algorithm \cite{Shor} cannot be applied to attack our scheme.

\section{Conclusion: advantages and limitations of the scheme}\label{limitations}

\subsection{Advantages} \hfill
%\medskip

\noindent {\bf 1.} A novel mathematical idea used for signature verification.
\medskip

\noindent {\bf 2.} Efficiency of the signature verification (about 0.3 sec on average).

\vskip 0.1 cm

\subsection{Limitations} \hfill
%\medskip

\noindent {\bf 1.} The main limitation is the size of the public key (about 15  Kbytes with suggested parameters).

The private key (the automorphism $\varphi$) is not too small either, about 1.5 Kbytes on average.
There is a trade-off between the size of $\varphi$ and its security. The size of $\varphi$ can be, in principle, reduced to just a few hundred monomials, but then security becomes a concern since some parts of $\varphi(x_i)$ may be possible to recover more or less by inspection of the public pairs $(P_i, \varphi(P_i))$.

The signature size is about 4 Kb on average, which is decent but not record-breaking.

%With currently suggested parameters, signature verification takes between 2-3 sec, which is not too bad, but the polynomial $\varphi(Q)$ (the main part of the signature) has about 10,000 monomials on average, so the signature is rather large. The size of $\varphi$ itself (i.e., the total number of monomials in all the polynomials $\varphi(x_i)$) is about 1500 monomials on average.

\medskip

\noindent {\bf 2.} Another limitation is that using non-deterministic methods, such as a Monte Carlo type method, may result in errors, more specifically in false negative or even false positive results of the signature verification, although so far, with suggested parameters, we did not detect any false negative or false positive results. (``False negative" means rejecting a valid signature.)

%Increasing the number of trials in the Monte Carlo method obviously reduces the probability of false negative as well as false positive results. %With 10,000 trials (instead of suggested 3000), we did not detect any false negative or false positive results.
%However, there is obviously a trade-off with the verification time.

\section*{Appendix}

Here we give a proof of Proposition \ref{positive}.
\medskip

\noindent {\bf  Proposition 1.}
Let $h=h(x_1,\ldots, x_n)$ be a polynomial from the set $\G$. Suppose $h$ does not depend on $x_k$.
Let $\alpha$ be the map that takes $x_k$ to $x_k + h - 2x_k \cdot h$ and fixes all other variables.
Then:
\medskip

\noindent {\bf (a)} $\alpha$ defines an automorphism of $B(K)$, the factor algebra of the algebra $\Z[x_1, \ldots, x_n]$ by the ideal generated by all polynomials of the form $(x_i^2-x_i)$, $i=1, \ldots, n$. Denote this automorphism also by $\alpha$.
\medskip

\noindent {\bf (b)} The group of automorphisms of $B(K)$ is generated by all automorphisms as in part (a) and is isomorphic to the group of permutations of the vertices of the $n$-dimensional Boolean cube.
\medskip

\noindent {\bf (c)} For any polynomial $P$ from $\Z[x_1, \ldots, x_n]$, the number of positive values of $P$ on Boolean tuples $(x_1, \ldots, x_{n})$ equals that of $\alpha(P)$.

\begin{proof}
{\bf (a)} Let $B^n$ denote the Boolean $n$-cube, i.e., the $n$-dimensional cube whose vertices are Boolean $n$-tuples. The map $\alpha$ leaves the set of vertices of $B^n$ invariant. Indeed, $\alpha$ fixes all $x_i$ except $x_k$, and it is straightforward to see that if $x_k=0$, then $\alpha(x_k) = h(x_1,\ldots, x_n)$, and if $x_k=1$, then $\alpha(x_k) = 1 - h(x_1,\ldots, x_n)$. Since on any Boolean $n$-tuple $(x_1, \ldots, x_n)$, one has $h(x_1, \ldots, x_n)=$  0 or 1 (see Section \ref{G}), we see that $\alpha$ is a bijection of the set of vertices of $B^n$ onto itself.

Next, observe that for any polynomial $h$ from the set $\G$, one has $h^2=h$ modulo the ideal generated by all polynomials of the form $(x_i^2-x_i)$; this easily follows from the inductive procedure of constructing polynomials $h$, see Section \ref{G}. Therefore, $\alpha$ leaves the ideal generated by all $(x_i^2-x_i)$ invariant since $\alpha$ takes $x_i$ to $x_i + h - 2x_i \cdot h$,  and then $\alpha(x_i^2-x_i) = (x_i + h - 2x_i \cdot h)^2 - (x_i + h - 2x_i \cdot h) = (x_i^2-x_i) + (h^2-h) + 2x_ih -4x_i^2h -4x_ih^2 + 4x_i^2h^2 +2x_ih = (x_i^2-x_i) + (h^2-h) + 4h(x_i-x_i^2) + 4h^2(x_i^2-x_i)$.
\medskip

\noindent {\bf (b)} Consider the automorphism $\alpha$ again. Fix a particular Boolean $n$-tuple $(x_1,\ldots, x_n)$. Suppose that $h(x_1,\ldots, x_n)=1$. Suppose $x_k=0$ in this tuple.  Then $\alpha$ takes this tuple to the tuple where all $x_i$, except $x_k$, are the same as before, and $x_k=1$, i.e., just one of the coordinates in the tuple was flipped. Therefore, an appropriate composition of different $\alpha$ (with different $x_k$) can map any given Boolean $n$-tuple to any other Boolean $n$-tuple.

\medskip

\noindent {\bf (c)} This follows immediately from the argument in the proof of part (a).
More specifically, since the set of vertices of $B^n$ is invariant under $\alpha$, there is a bijection between the sets of values of $P$ and $\alpha(P)$ on Boolean $n$-tuples.

\end{proof}

%\subsection{Applying $\varphi$ to the polynomial $Q = Q(x_1,...,x_{n+1})$}
%\medskip
%
%To apply $\varphi$ to the polynomial $Q$, plug in
%$\varphi(x_i)$ for each $x_i, ~i = 1,...,9$ in the polynomial $Q$. After that, take $x_{n+1}$ to $x_{n+1} + r(x_1,...,x_n)$.

\end{document}